\documentclass[12pt]{article}

\usepackage[bf,sf]{titlesec}
\usepackage[utf8]{inputenc}

\usepackage{subfigure}
\usepackage[round,authoryear]{natbib}

\usepackage{bm}
\usepackage{amsmath}

\usepackage{boxedminipage}

\usepackage{listings}

\usepackage{ifpdf}

\usepackage[T1]{fontenc}
\usepackage[bf,sf]{titlesec}	
\usepackage[Sonny]{fncychap} 
  \ChNameVar{\Large\sf\bfseries}
  \ChNumVar{\Huge\bfseries}
  \ChTitleVar{\Large\sf\bfseries}
\usepackage{fancyhdr}   
\usepackage{makeidx}

\ifpdf
\usepackage[pdftex]{graphicx}
\else
\usepackage{graphicx}
\fi
\usepackage{rotating}
\usepackage[includemp=false, marginparwidth=1.8cm, hmargin=3.5cm, vmargin=3cm]{geometry}
\ifnum\pdfoutput>0			
\usepackage[pdftex, 			
pagebackref = false, 			
bookmarks = true, 			
bookmarksnumbered = false, 	
hyperindex = true,			
linktocpage = true,			
pdfpagemode = UseNone, 		
pdfstartview = Fit, 			
pdfpagelayout = OneColumn,
colorlinks = true, 			
urlcolor = blue, 				
citecolor = blue,
pdfborder = {0 0 0} 			
]{hyperref} 					
\fi

\title{Discussion of ``Riemann manifold Langevin and Hamiltonian Monte Carlo methods'' by M. Girolami and B. Calderhead}
\author{Luke Bornn\footnote{University of British Columbia, Department of Statistics, \url{l.bornn@stat.ubc.ca}}, Julien Cornebise\footnote{University of British Columbia, Department of Statistics and Department of Computer Science, \url{cornebis@cs.ubc.ca}}, Gareth W. Peters\footnote{University of New South Wales, School of Mathematics and Statistics, \url{garethpeters@unsw.edu.au}}}
\date{\today}

\begin{document}

\ifpdf
\DeclareGraphicsExtensions{.png, .pdf, .jpg, .tif}
\else
\DeclareGraphicsExtensions{.eps, .jpg}
\fi

\maketitle

\section*{Introduction}

This technical report is the union of two contributions to the
discussion of the Read Paper 
\emph{Riemann manifold Langevin and Hamiltonian Monte Carlo methods} \citep*{CalderheadGirolami2010},
presented in front of the Royal Statistical Society on October 13\textsuperscript{th} 2010 
and to appear in the Journal of the Royal Statistical Society Series B.

The first comment establishes a parallel and possible interactions with Adaptive Monte Carlo methods. The second comment exposes a detailed study of Riemannian Manifold Hamiltonian Monte Carlo (RMHMC) for a weakly identifiable model presenting a strong ridge in its geometry.

\section{On Adaptive Monte Carlo \small{-- J. Cornebise and G. Peters}}

The utility of RMMALA and RMHMC methodology is its ability to adapt Markov
chain proposals to the current state. Many articles design Adaptive Monte Carlo
(MC) algorithms to learn efficient MCMC proposals, such as the special case of
controlled MCMC, \citet{Haario2001adaptive} which utilizes a historically
estimated global covariance for a Random Walk Metropolis Hastings (RWMH)
algorithm. Similarly, \citet{Atchade2006adaptive} devised global adaptation in
MALA. Surveys are provided in \cite{atchade2009adaptive},
\cite{andrieu2008tutorial} and \cite{RobertsRosenthal2009examples}.

When the proposal remains essentially unchanged regardless of the current
Markov chain state, performance may be poor if the shape of the target
distribution varies widely over the parameter space. A typical illustration
involves the ``banana-shaped'' warped Gaussian (see comments by Cornebise and
Bornn), ironically originally utilized to illustrate the strength of early
Adaptive MC \citep[Figure 1]{Haario1999}. This algorithm learned local geometry
by estimating the covariance matrix based on a sliding window of past states.
However, \citet[Appendix A]{Haario2001adaptive} showed it could exhibit strong
bias, perhaps connected to requirements of ``diminishing adaptation'' as
studied in \citet{AndrieuMoulines2006ergodicity}. Recent locally adaptive
algorithms satisfy this condition, e.g. \emph{State-dependent proposal
scalings} \citep[Section 3.4]{Rosenthal2010optimal} fits a parametric family to
the covariance as a function of the state, or the parameterized parameter space
approach of \emph{Regional Adaptive Metropolis Algorithms} \citep[Section
5]{RobertsRosenthal2009examples}.

Riemannian approaches provide strong rationale for parameterizing the proposal
covariance as a function of the state -- without learning, when the FIM (or
observed FIM) can be computed or estimated, (see comment by Doucet and Jacob).
With unknown FIM, or to learn the optimal step size, it would be interesting to
combine Riemannian Monte Carlo with adaption. A first step could involve a
simplistic Riemann-inspired algorithm such as a centered RWMH via the
(observed) FIM as the proposal covariance \citep[Section 4.3.1][as used
in]{MarinRobert2007bayesiancore} -- equivalent to one step of RMMALA without
drift.

An additional use of Riemannian MC could be within the MCMC step of Particle
MCMC \citep{AndrieuDoucetHolenstein2010}, where adaption was highly
advantageous in the AdPMCMC algorithms of \cite{peters2010ecological}.

Another interesting extension involves considering the stochastic approximation
alternative approach, based on a curvature updating criterion of \cite[Equation
10]{Okabayashi2010}, for an adaptive line search. This was proposed as an
alternative to MCMC-MLE of \cite{geyer1992markov} for complex dependence
structures in exponential family models. In particular comparing properties of
this curvature based condition based on local gradient information with
adaptive RMMALA and RMHMC versions of the MCMC-MLE algorithms would be
instructive.

Additionally, one may consider how to extend Riemannian MC to trans-dimensional
MC such as reversible jump \citep{RichardsonGreen1997}, for which adaptive
extensions are rare \citep[Section 4.2]{GreenHastie2009reversible}. We wonder
how a geometric approach may be extended to efficiently explore disjoint unions
of model subspaces as in \citet{NevatPetersYuan2009contour}.

Finally, an open question to the community: could such
geometric tools be utilized in Approximate Bayesian Computation
\citep{BeaumontCornuetMarinRobert2009}, e.g. to design the distance metric
between summary statistics?

\section{RMHMC for unidentifiable models \small{-- J. Cornebise and L. Bornn}}

In this comment we show how the proposed RMHMC method can be particularly useful
in the case of strong geometric features such as ridges commonly occurring in
nonidentifiable models.  While it has been suggested to use tempering or adaptive methods to handle these ridges \citep[][e.g.]{Neal2001, Haario2001adaptive}, they remain a celebrated challenge for new Monte
Carlo methods \citep{Cornuet2009AMIS}.  We suspect that RMHMC, by exploiting the geometry of the surface to help make intelligent moves along the ridge, is a brilliant advance for nonidentifiability sampling issues.

\begin{figure}
	\centering
	\includegraphics[width=\textwidth]{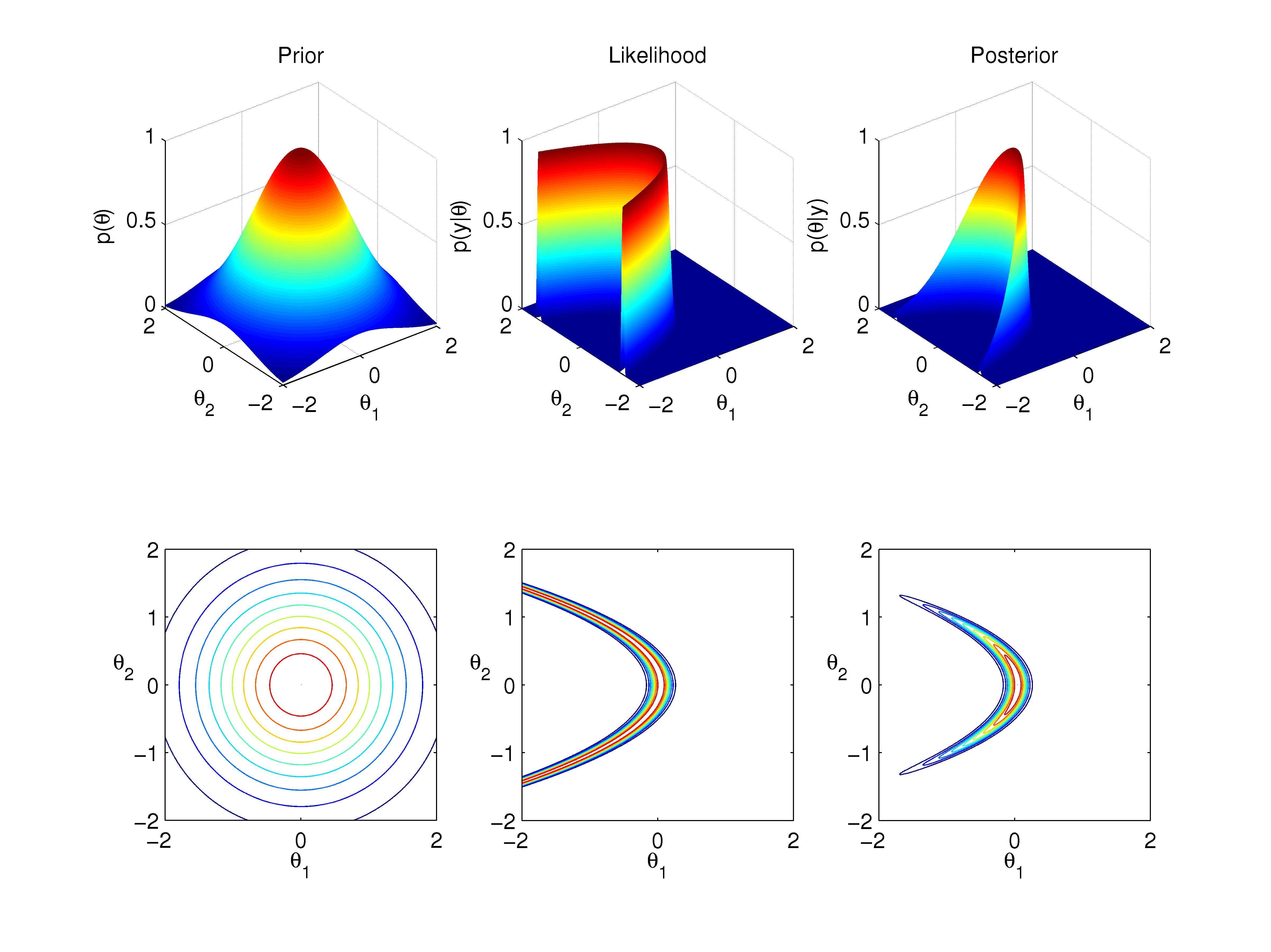}
	\caption{Prior, likelihood, and posterior for the warped bivariate Gaussian with 
	$n = 100$ values generated from the likelihood with parameter settings $\sigma_{\theta}=\sigma_y=1$.  As the sample size increases and the prior becomes more diffuse, the posterior becomes less identifiable and the ridge in the posterior becomes stronger.}
	\label{densities}
\end{figure}

\begin{figure}
  \centering
	\subfigure[RMHMC]{
      \includegraphics[width=0.45\textwidth]{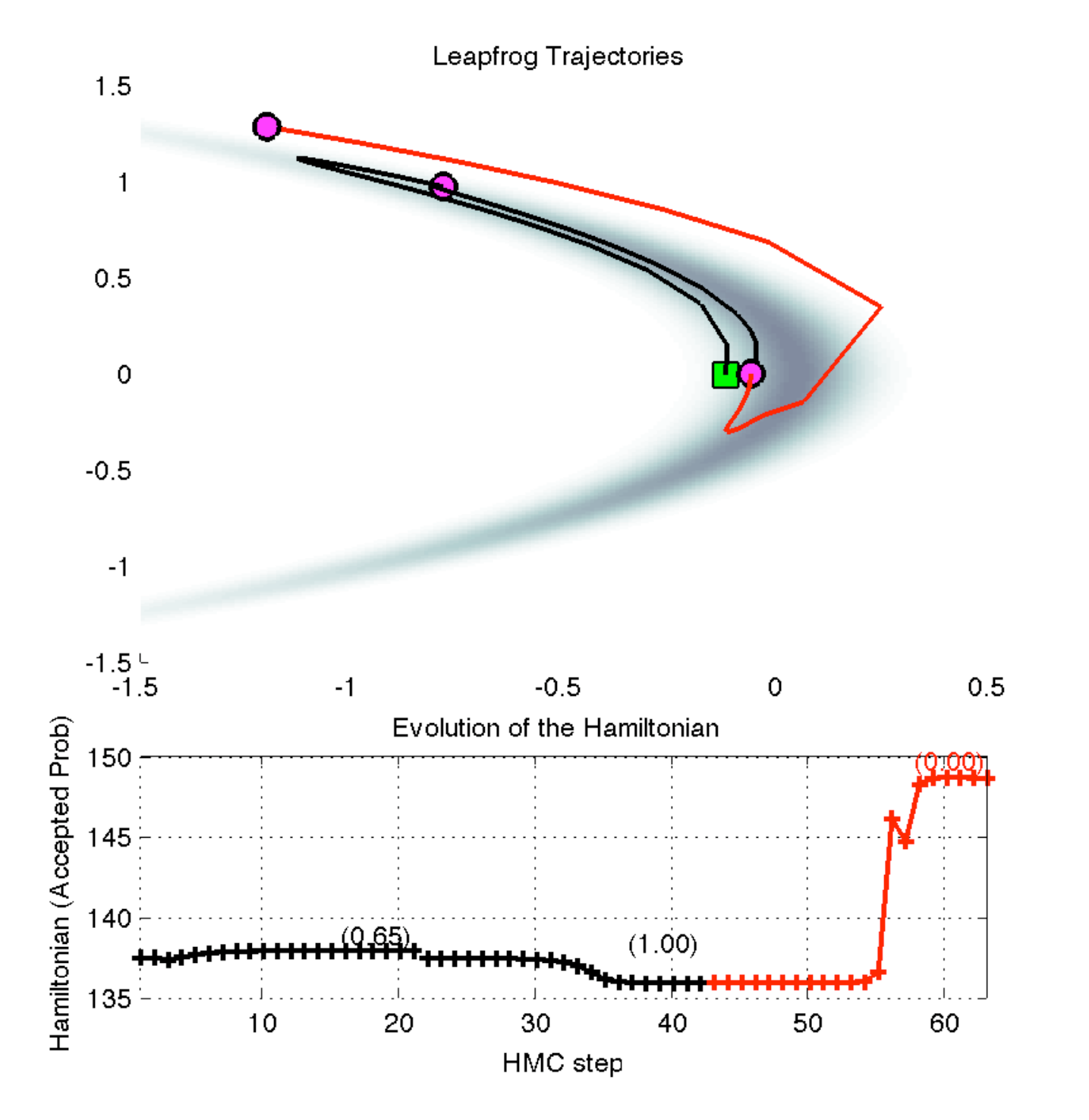}
	}
	\subfigure[HMC]{
      \includegraphics[width=0.45\textwidth]{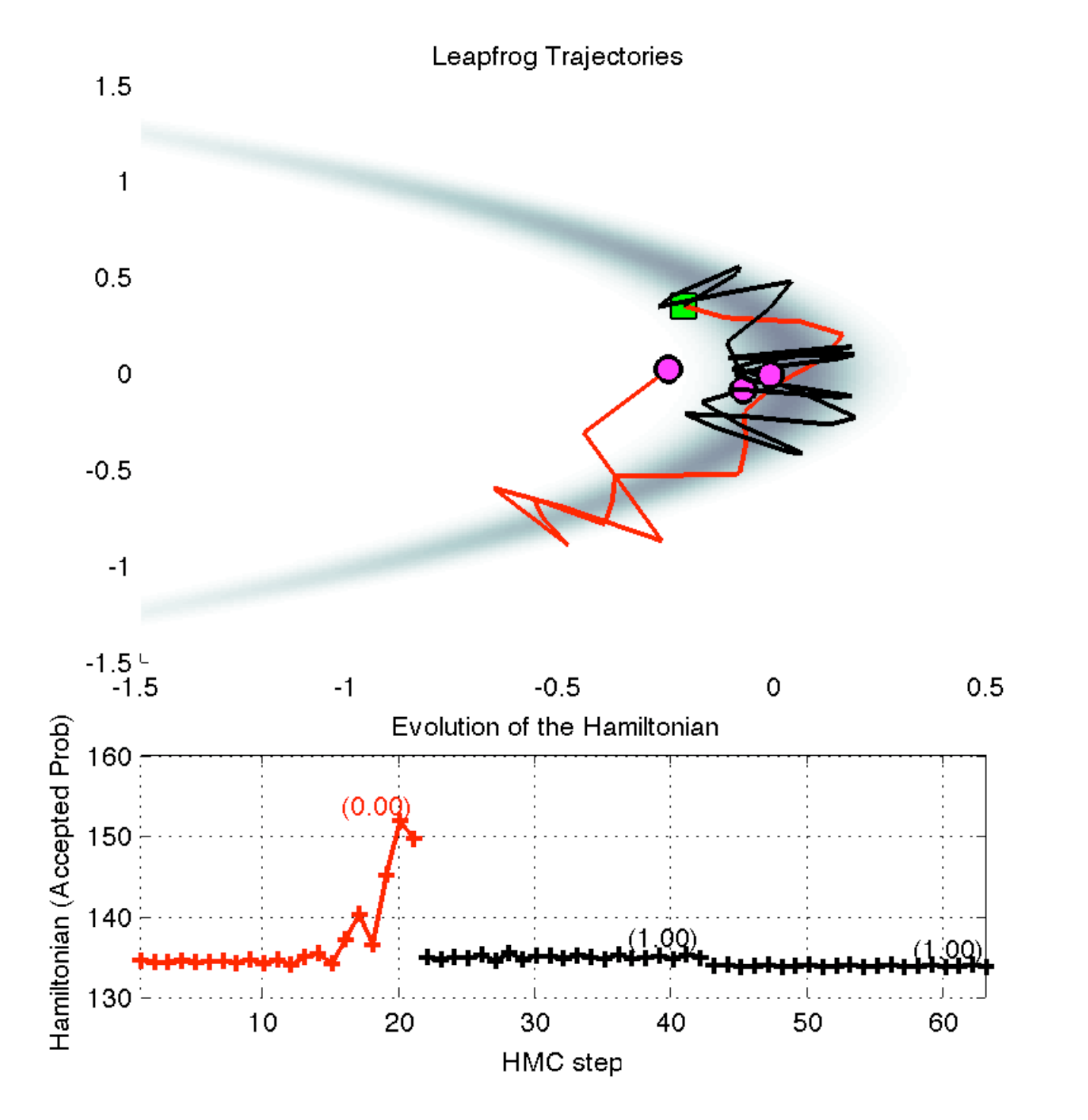}
	}
	\caption{Three typical consecutive trajectories of 20 leapfrog steps each, with step size of $0.1$,
	for the RMHMC and HMC algorithm, chosen to highlight two acceptances (black) and one rejection (red), representative of the approximately $65\%$ acceptance ratio for both HMC and RMHMC.
  We see that RMHMC is able to track the contours of the density and reach to
  the furthest tails of the ridge, adapting to the local geometry, whereas the spherical moves of HMC oscillate back and forth across the ridge.}
	\label{3Traj}
\end{figure}

Consider observations $y_1,\dots, y_n \sim \mathcal{N}(\theta_1 + \theta_2^2,
\sigma_y^2)$.  
The parameters $\theta_1$ and $\theta_2$ are non-identifiable without any
additional information beyond the observations: any values such that $\theta_1
+ \theta_2^2 = c$ for some constant $c$ explain the data
equally well. By imposing a prior distribution, $\theta_1, \theta_2 \sim \mathcal{N}(0,
\sigma_{\theta}^2)$, we create weak identifiability, namely decreased posterior
probability for $c$ far from zero.  Figure \ref{densities} shows the prior, likelihood, and ridge-like posterior for the model.
For this problem, we have \begin{align*}
	\bm{G}(\bm \theta) = \left( \begin{array}{cc}
	\displaystyle \frac{n}{\sigma_y^2} + \frac{1}{\sigma_{\theta}^2}
	& \displaystyle \frac{2n\theta_2}{\sigma_y^2} 
	\\[1em] \displaystyle \frac{2n\theta_2}{\sigma_y^2} 
	& \displaystyle \frac{4n\theta_2^2}{\sigma_y^2} + \frac{1}{\sigma_{\theta}^2} 
	\end{array} \right).
\end{align*}
Figure \ref{3Traj} compares typical trajectories of both HMC and RMHMC, demonstrating the ability of RMHMC to follow the full length of the ridge.

HMC and RMHMC also differ in sensitivity to the step size. 
As described by \citet{Neal2010}, HMC suffers from the presence of a critical step size above which the error explodes, accumulating at each leapfrog step.
In contrast, RMHMC occasionally exhibits a sudden jump in the Hamiltonian at one specific leapfrog step,
followed by well-behaving steps (as seen in Figure 2.(a)). This is due
to the possible divergence of the fixed point iterations (FPI) in the generalized leapfrog equations
	\newcommand{\p}{\mathbf{p}}
	\newcommand{\he}{\frac{\varepsilon}{2}}
\begin{equation}
	\p\left( \tau + \he \right) = \p(\tau) - \he \nabla_{\bm \theta} H \left\{
	\bm\theta\left( \tau \right), \p\left( \tau + \he \right) \right\}
	\tag{16}
\end{equation}
for given momentum $\p(\tau)$, parameter $\bm\theta(\tau)$ and step size $\varepsilon$. Figure \ref{validprob} shows the probability of (16) having a solution
$\p(\varepsilon/2)$ as a
function of $\bm \theta(0)$, and of the derivative at the fixed point being ``small enough'' for the FPI to converge; the well-known sufficient theoretical threshold on the derivative \citep[][see e.g.]{Fletcher1987} is $1$, but we conservatively chose $1.2$ based on typical successful runs.
\begin{figure}
	\centering
	\subfigure[Probability of existence of fixed point]{
		\includegraphics[width=0.7\textwidth]{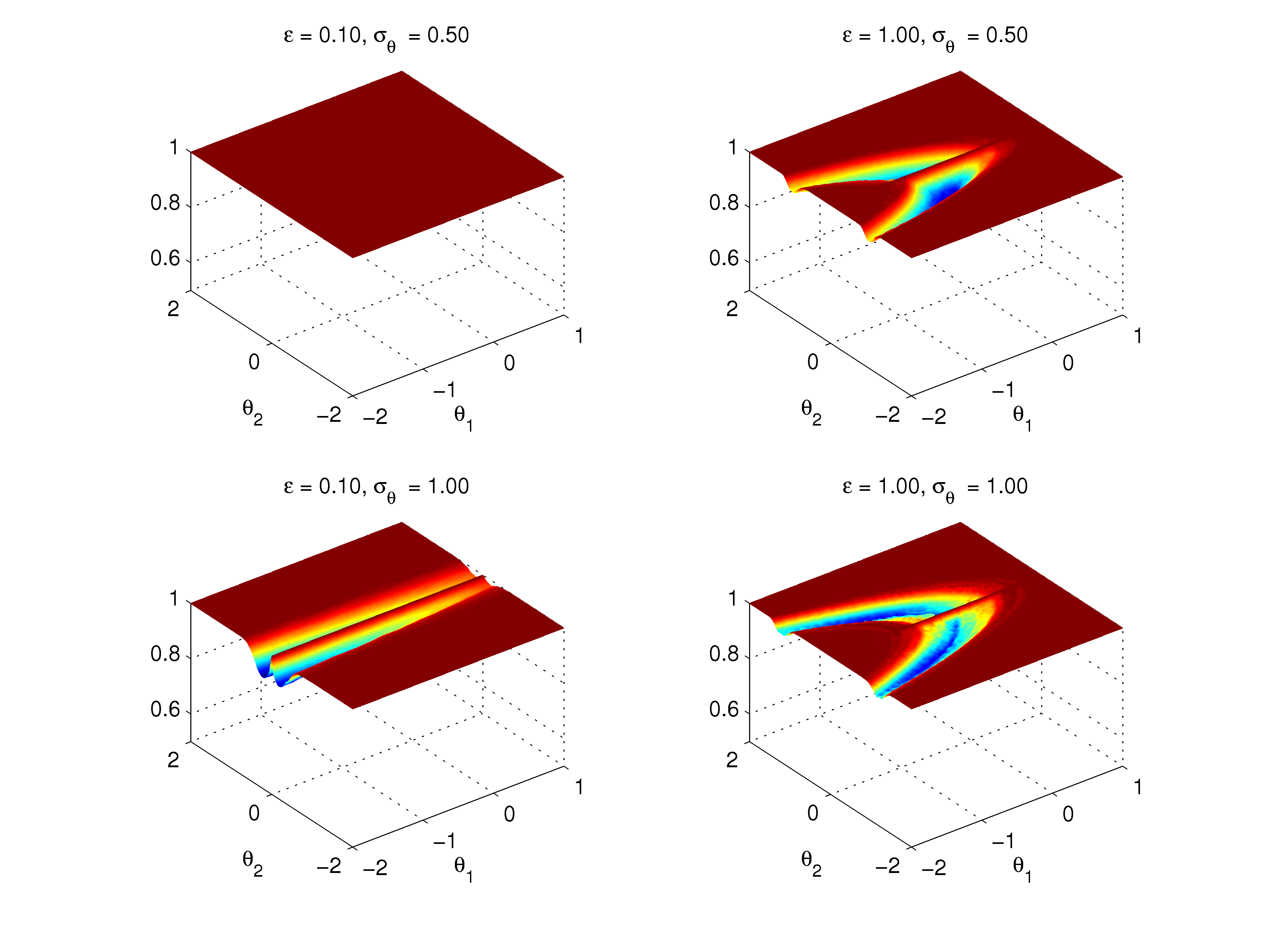}
	}
	\subfigure[Probability of existence of fixed point and of a derivative lower than 1.2]{
		\includegraphics[width=0.7\textwidth]{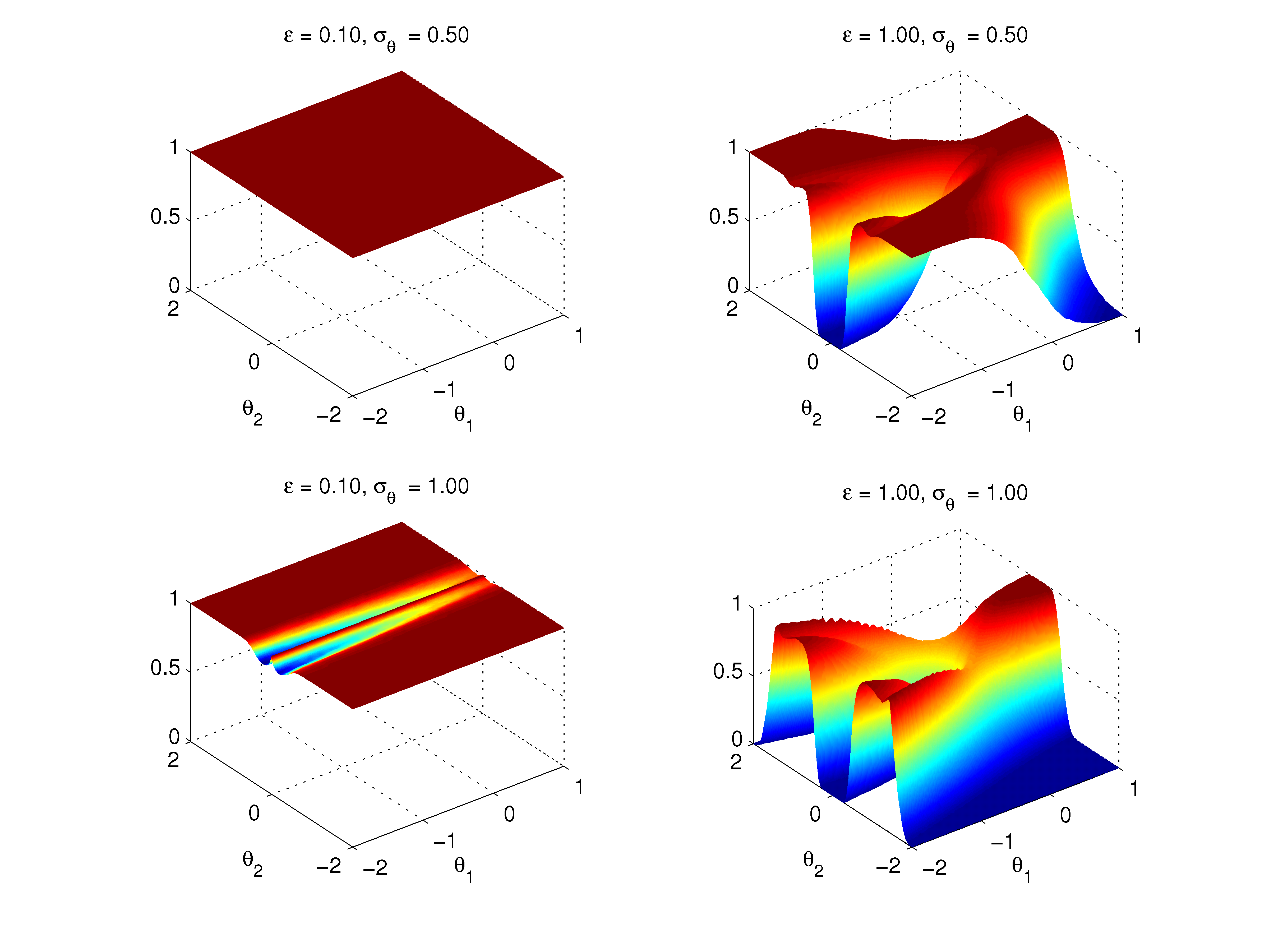}
	}
	\caption{Probability of a single iteration of the generalized leapfrog (a) admitting
	a fixed point $\p\left( \he \right)$, (b) admitting a fixed point and having a small enough derivative at the fixed point for the 
	FPI to converge. Both plotted as a function of starting point $\bm\theta(0)$.  Shown for two
	step sizes $\varepsilon \in \{0.1, 1.0\}$, and two prior distribution standard deviations
	$\sigma_{\theta} \in \{0.5, 1.0\}$, i.e. varying levels of identifiability. The region of
	stability for FPI becomes much smaller as
	the step size increases and as identifiability decreases, even creating regions with null probability of convergence.}
	\label{validprob}
\end{figure}
When the finite number of FPI diverges, 
the Hamiltonian explodes; however, subsequent steps may still admit a fixed point, and hence behave normally. Unsurprisingly, this behavior is much more likely to occur for larger step sizes.

While the regions of low probability 
can strongly decrease the mixing of the algorithm, they do not affect the
theoretical convergence ensured by the rejection step. 
Far from being a downside, understanding this behavior can bring much practical insight when choosing
the step size -- possibly adapting it on-the-fly, when RMHMC already
provides a clever way to adaptively devise the direction of the moves.

\begin{small}
\bibliographystyle{plainnat}
\bibliography{biblio}

\begin{thebibliography}{22}
\providecommand{\natexlab}[1]{#1}
\providecommand{\url}[1]{\texttt{#1}}
\expandafter\ifx\csname urlstyle\endcsname\relax
  \providecommand{\doi}[1]{doi: #1}\else
  \providecommand{\doi}{doi: \begingroup \urlstyle{rm}\Url}\fi

\bibitem[Andrieu and Moulines(2006)]{AndrieuMoulines2006ergodicity}
C.~Andrieu and {\'E}.~Moulines.
\newblock {On the ergodicity properties of some adaptive MCMC algorithms}.
\newblock \emph{The Annals of Applied Probability}, 16\penalty0 (3):\penalty0
  1462--1505, 2006.

\bibitem[Andrieu and Thoms(2008)]{andrieu2008tutorial}
C.~Andrieu and J.~Thoms.
\newblock {A tutorial on adaptive MCMC}.
\newblock \emph{Statistics and Computing}, 18\penalty0 (4):\penalty0 343--373,
  2008.

\bibitem[Andrieu et~al.(2010)Andrieu, Doucet, and
  Holenstein]{AndrieuDoucetHolenstein2010}
C.~Andrieu, A.~Doucet, and R.~Holenstein.
\newblock Particle {M}arkov chain {M}onte {C}arlo methods.
\newblock \emph{J. R. Statis. Soc. B}, 72\penalty0 (3):\penalty0 269--342,
  2010.

\bibitem[Atchad{\'e} et~al.(2009)Atchad{\'e}, Fort, Moulines, and
  Priouret]{atchade2009adaptive}
Y.~Atchad{\'e}, G.~Fort, E.~Moulines, and P.~Priouret.
\newblock {Adaptive Markov Chain Monte Carlo: Theory and Methods}.
\newblock Technical report, 2009.

\bibitem[Atchad{\'e}(2006)]{Atchade2006adaptive}
Y.F. Atchad{\'e}.
\newblock {An adaptive version for the Metropolis adjusted Langevin algorithm
  with a truncated drift}.
\newblock \emph{Methodology and Computing in Applied Probability}, 8\penalty0
  (2):\penalty0 235--254, 2006.

\bibitem[Beaumont et~al.(2009)Beaumont, Cornuet, Marin, and
  Robert]{BeaumontCornuetMarinRobert2009}
M.A. Beaumont, J.M. Cornuet, J.M. Marin, and C.P. Robert.
\newblock Adaptive approximate bayesian computation.
\newblock \emph{Biometrika}, 96\penalty0 (4):\penalty0 983--990, 2009.

\bibitem[Calderhead and Girolami(2010)]{CalderheadGirolami2010}
B.~Calderhead and M.~Girolami.
\newblock {Riemann manifold Langevin and Hamiltonian Monte Carlo methods (with
  discussion)}.
\newblock \emph{Journal of the Royal Statistical Society: Series B}, to appear,
  2010.

\bibitem[Cornuet et~al.(2009)Cornuet, Marin, Mira, and Robert]{Cornuet2009AMIS}
J.M. Cornuet, J.M. Marin, A.~Mira, and C.~Robert.
\newblock {Adaptive multiple importance sampling}.
\newblock Technical report, ArXiV, 2009.
\newblock URL \url{http://arxiv.org/abs/0907.1254}.

\bibitem[Fletcher(1987)]{Fletcher1987}
R.~Fletcher.
\newblock \emph{Practical Methods of Optimization, second edition}.
\newblock 1987.

\bibitem[Geyer(1991)]{geyer1992markov}
C.J. Geyer.
\newblock {Markov chain Monte Carlo maximum likelihood}.
\newblock In \emph{Computing {S}cience and {S}tatistics: {P}roc. 23rd {S}ymp.
  {I}nterface}, pages 156--163, 1991.

\bibitem[Green and Hastie(2009)]{GreenHastie2009reversible}
P.J. Green and D.I. Hastie.
\newblock Reversible jump {MCMC}.
\newblock Technical report, June 2009.

\bibitem[Haario et~al.(1999)Haario, Saksman, and Tamminen]{Haario1999}
H.~Haario, E.~Saksman, and J.~Tamminen.
\newblock Adaptive proposal distribution for random walk {M}etropolis
  algorithm.
\newblock \emph{Computational Statistics}, 14:\penalty0 375--395, 1999.

\bibitem[Haario et~al.(2001)Haario, Saksman, and Tamminen]{Haario2001adaptive}
H.~Haario, E.~Saksman, and J.~Tamminen.
\newblock {An adaptive Metropolis algorithm}.
\newblock \emph{Bernoulli}, 7\penalty0 (2):\penalty0 223--242, 2001.

\bibitem[Marin and Robert(2007)]{MarinRobert2007bayesiancore}
J.M. Marin and C.P. Robert.
\newblock \emph{{Bayesian core: a practical approach to computational Bayesian
  statistics}}.
\newblock Springer Verlag, 2007.

\bibitem[Neal(2001)]{Neal2001}
R.M. Neal.
\newblock {Annealed importance sampling}.
\newblock \emph{Statistics and Computing}, 11\penalty0 (2):\penalty0 125--139,
  2001.

\bibitem[Neal(2010)]{Neal2010}
R.M. Neal.
\newblock {MCMC using Hamiltonian dynamics}.
\newblock In S.~Brooks, A.~Gelman, G.~Jones, and X.L. Meng, editors,
  \emph{Handbook of {Markov Chain Monte Carlo}}. Chapman and Hall/CRC Press,
  2010.

\bibitem[Nevat et~al.(2009)Nevat, Peters, and Yuan]{NevatPetersYuan2009contour}
I.~Nevat, G.W. Peters, and J.~Yuan.
\newblock {Channel Estimation in OFDM Systems with Unknown Power Delay Profile
  using Trans-Dimensional MCMC via Stochastic Approximation}.
\newblock In \emph{Vehicular Technology Conference, 2009. VTC Spring 2009. IEEE
  69th}, pages 1--6. IEEE, 2009.

\bibitem[Okabayashi and Geyer(2010)]{Okabayashi2010}
S.~Okabayashi and C.~J. Geyer.
\newblock Long range search for maximum likelihood in exponential families.
\newblock Technical report, 2010.

\bibitem[Peters et~al.(2010)Peters, Hosack, and Hayes]{peters2010ecological}
G.W. Peters, G.R. Hosack, and K.R. Hayes.
\newblock {Ecological non-linear state space model selection via adaptive
  particle Markov chain Monte Carlo (AdPMCMC)}.
\newblock Technical report, 2010.

\bibitem[Richardson and Green(1997)]{RichardsonGreen1997}
S.~Richardson and P.J. Green.
\newblock {On Bayesian analysis of mixtures with an unknown number of
  components (with discussion)}.
\newblock \emph{Journal of the Royal Statistical Society: Series B},
  59\penalty0 (4):\penalty0 731--792, 1997.

\bibitem[Roberts and Rosenthal(2009)]{RobertsRosenthal2009examples}
G.O. Roberts and J.S. Rosenthal.
\newblock {Examples of adaptive MCMC}.
\newblock \emph{Journal of Computational and Graphical Statistics}, 18\penalty0
  (2):\penalty0 349--367, 2009.

\bibitem[Rosenthal(2010)]{Rosenthal2010optimal}
J.S. Rosenthal.
\newblock {Optimal proposal distributions and adaptive MCMC}.
\newblock In S.~Brooks, A.~Gelman, G.~Jones, and X.L. Meng, editors,
  \emph{Handbook of {Markov Chain Monte Carlo}}. Chapman and Hall/CRC Press,
  2010.

\end{thebibliography}

\end{small}
\end{document}